\documentclass[a4paper,fleqn,usenatbib]{mnras}

\usepackage{newtxtext,newtxmath}

\usepackage[T1]{fontenc}
\usepackage{ae,aecompl}

\usepackage{graphicx}   
\usepackage{amsmath}    
\usepackage{amssymb}    
\usepackage{amsfonts}
\usepackage{bm,color}

\newcommand{\der}[2]{\frac{\partial #1}{\partial #2}}
\newcommand{\dder}[3]{\frac{\partial^2 #1}{\partial #2\partial #3}}

\newcommand{\w}[1]{\bm{#1}}
\newcommand{\vv}[1]{\vec{\w{#1}}}

\newcommand{\be}{\begin{equation}}
\newcommand{\ee}{\end{equation}}
\newcommand{\bea}{\begin{eqnarray}}
\newcommand{\eea}{\end{eqnarray}}

\newcommand{\Liec}[1]{{\mathcal{L}}_{\vv{#1}}\,}

\title[Rotating Neutron Stars in $F(R)$ Gravity with Axions]{Rotating Neutron
Stars in $F(R)$ Gravity with Axions}
\author[Astashenok \& Odintsov]{Artyom V. Astashenok$^{1}$, Sergey D.
Odintsov$^{2,3}$\\
\small $^{1}$Immanuel Kant Baltic Federal University\\
\small Department of Physics, Technology and IT\\
\small 236041 Kaliningrad, Russia, Nevskogo str. 14\\
\small $^{2}$Institut de Ci\'{e}ncies de l'Espai, ICE/CSIC-IEEC, Campus UAB,
Carrer de Can Magrans s/n, 08193 Bellaterra (Barcelona), Spain \\
\small $^{3}$Instituci\'{o} Catalana de Recerca i Estudis
Avan\c{c}ats (ICREA), Barcelona, Spain}

\begin{document}
\label{firstpage}
\pagerange{\pageref{firstpage}--\pageref{lastpage}} \maketitle

\maketitle

\begin{abstract}

We investigate equilibrium configurations of uniformly rotating
neutron stars in $R^2$ gravity with axion scalar field for GM1
equation of state (EoS) for nuclear matter. The mass-radius
diagram, mass-central energy density are presented for some
frequencies in comparison with static stars. We also compute
equatorial and polar radii and moment of inertia for stars. For
axion field $\phi$  the coupling in the form $\sim R^2\phi$ is
assumed. Several interesting results follow from our
consideration. Maximal possible star mass with given EoS increases
due to the contribution of coupling term. We discovered the
possibility to increase maximal frequency of the rotation in
comparison with General Relativity. As a consequence the lower
bound on mass of the fast rotating stars decreases. For frequency
$f=700$ Hz neutron  stars with masses $\sim M_\odot$ can exist for
some choice of parameters (in General Relativity for same EoS this
limit is around $1.2 M_{\odot}$). Another feature of our solutions
is relatively small increase of stars radii for high frequencies
in comparison with static case. Thus, eventually the  new class of
neutron stars in $R^2$ gravity with axions is discovered namely
fast rotating compact stars with intermediate masses.
\end{abstract}


\section{Introduction}

Consistent description of rotating neutron stars is one of the
most interesting problems in modern astrophysics. From theoretical
viewpoint it can be considered as a test for General Relativity
and our models for strong interactions at very high densities
($10^{15}-10^{16}$ g/cm$^{3}$).

Since the pioneer work of Oppenheimer and Volkoff
(\citealp{Oppenheimer}) our knowledge about neutron stars has been
considerably extended.  Mass, radius and other parameters of
relativistic stars depend from the equation of state (EoS) chosen
for dense matter. Tens equations of state for description of
neutron star matter were proposed over the years
\citealp{Rezzolla}. J. Antoniadis and colleagues confirmed the
existence of massive neutron stars with $M>2M_\odot$ measuring the
mass of PSR J0348+0432  with help of white dwarf spectroscopy
(\citealp{Antoniadis}; \citealp{Antoniadis-2}). This limit
constrains dramatically the stiffness of nuclear EoS: so-called
hyperon puzzle takes place (see recent review of \citealp{Tolos}
and reference therein).

Slow and fast rotating neutron stars  have been investigated
mainly in frames of General Relativity. Slow-rotation
approximation and its second order was firstly investigated by
\citealp{Hartle}, \citealp{Hartle-2}. Consideration of slow
rotation regime for uniformly rotating stars helps to establish
universal relations between quadrupole moment, moment of inertia,
and Love number of neutron stars (\citealp{Pappas},
\citealp{Yagi}, \citealp{Yagi-2}, \citealp{Yagi-3}).

Numerical procedures for constructing of equilibrium stellar
configurations for the case of fast rotation were developed in many
works (see for example \citealp{Bonazzola-93},
\citealp{Shapiro-94}, \citealp{Stergioulas-95}, and used in some
papers (see e.g. \citealp{Chakrabarti} and references therein) for
obtaining similar EoS independent relations in full rotation
regime. Differential rotating of neutron stars with polytropic EoS
were studied in \citealp{Rezzola-2} with using magnetohydrodynamic
simulations (\citealp{Rezzola-3}). Interesting results concerning
rotating neutron stars with magnetic fields were obtained by
\citealp{Rezzola-4}, \citealp{Rezzola-5}, \citealp{Rezzola-6}.

The gravitational field in neutron stars is extremely strong and
therefore the question about possible deviations from General
Relativity appears. In principle in frames of modified gravity one
can obtain new branches of compact stars and its possible
observation can help to discriminate between General Relativity and
its counterparts.

Models of modified gravity are also motivated by cosmological
background. Problem of dark energy which lead to accelerated
expansion of universe (\citealp{Riess-1}; \citealp{Perlmutter};
\citealp{Riess-2}) is usually treated in context of $\Lambda $CDM
model. According to this model dark energy is nothing else than
vacuum energy or cosmological constant with density consisting of around 70\%
of the all energy density in the universe. Usual
baryon matter gives only 4 \%. The rest part is so-called dark
matter. This is another unresolved puzzle of modern astrophysics
and cosmology.

Particle nature of dark matter is not questioned. A lot of
astrophysical data support this viewpoint. As an example one should
  mention data about collision of galaxies in the Bullet Cluster
and {cluster MACSJ0025} (\citealp{Markevitch}; \citealp{Clowe};
\citealp{Robertson}; \citealp{Bradac})). There are two main
candidates on the role of dark matter: weakly interacted massive
particles (WIMPs) and axions (\citealp{Sakharov};
\citealp{Sakharov-2}; \citealp{Sakharov-3}; \citealp{Marsh};
\citealp{Marsh-2}; \citealp{Oikonomou}; \citealp{Cicoli};
\citealp{Fukunaga}; \citealp{Caputo}). Direct experiments for
WIMPs detection give negative results ({see} \citealp{Ahmed};
\citealp{Davis}; \citealp{Davis2}; \citealp{Roszkowski};
\citealp{Schumann}). Otherwise some indications in favor of
existence of axions take place (\citealp{ADMX};
\citealp{ABRACADABRA}; \citealp{Quellet}; \citealp{Safdi};
\citealp{Avignone}; \citealp{Caputo-2}; \citealp{Caputo-3};
\citealp{AX-1}; \citealp{Rozner}). The possibility of axions
detection is based on axion-photon interaction in the presence of
magnetic fields (\citealp{AX-3}; \citealp{AX-4}; \citealp{AX-5}).
According to theoretical estimations axion mass is very small but
can lie in the wide limits $\sim 10^{-12}-10^{-3}$ eV. In the context
of high energy astrophysics axions can affect on process of the
cooling of neutron stars (see \citealp{Sedrakian},
\citealp{Sedrakian-2}, \citealp{Sedrakian-3}). They cause
instabilities in neutron star magnetosphere (\citealp{McDonald})
or even can mediate strong forces between neutron stars in binary
system (see \citealp{Anson}).

Alternative description of accelerated cosmological expansion is
proposed in various models of modified gravity
(\citealp{Capozziello1}; \citealp{Odintsov1}; \citealp{Turner}).
One should note also possibility of the unified description of
cosmological evolution including early inflation, matter and
radiation dominance era in $f(R)$ theory (\citealp{Nojiri-5};
\citealp{Nojiri-4}).

The interesting model of $f(R)$ gravity with axion dark matter was
proposed recently by \citealp{Oikonomou-19}. Using simple
misalignment model (\citealp{AX-2}) for axion field and $R^2$
gravity with non-minimal coupling with axion field it could describe early
inflation and dark energy era within the same model.

Compact non-rotating stars in simple models of $f(R)$ gravity were
extensively investigated in many works (for recent review of
compact star models in modified theories of gravity see
\citealp{Olmo} and references therein). Perturbative approach at
which scalar curvature is assumed as $R \sim T$ ($T$ is the trace
of energy-momentum tensor) was studied by \citealp{Arapoglu2011};
\citealp{Alavirad2013}; \citealp{Astashenok2013};
\citealp{Astashenok2014}; \citealp{Cheoun2013};
\citealp{Astashenok2015}. Self-consistent models of quark and
neutron stars in $R^2$ gravity  were considered
  in ref. \citealp{Astashenok2015-2};
\citealp{Astashenok2017}; \citealp{Astashenok2018}. Some
interesting results were obtained. In General Relativity the
solution outside the neutron star coincides with Schwarzschild
solution around the star with some mass $M$. But in $R^2$ gravity
the solution near the conventional surface of star (where pressure
of matter drops to zero) is not Schwarzschild one because scalar
curvature $R\neq 0$ outside the star. Scalar curvature quickly
drops and from some distance one can assume that $R=0$ and
therefore we have a solution corresponding to Schwarzschild
solution with some mass $M_s$ and $M_s$ is not equal to mass
confined by star surface. From the viewpoint of distant observer
mass $M_s$ is gravitational mass of neutron star. One should
mention that possible observable consequences appear only if the
contribution of $R^2$-term  is sufficiently large. It is
interesting to consider models of $R^2$ gravity in which
contribution of quadratic term is driven by some scalar field
$\phi$. To construct such model it is sufficient to add the
coupling between curvature and scalar field in simple form
$R^2\phi$. Assuming for scalar field the solution of ``core type''
inside star one can expect that coupling term can strengthen the
contribution of $R^2$-term. Outside the star the scalar field and
scalar curvature quickly damp (of course, in comparison with the
corresponding values inside the star). The radius of core should
be around of Compton wavelength $\lambda_c$ for scalar field
particles. For example for axion with mass in range $m_{a}\sim
10^{-11}-10^{-10}$ eV $\lambda_c \sim
  10-10^2$ km. This scale is comparable with characteristic size of
neutron stars.

Static configurations in this model of gravity have been
considered recently by us (see \citealp{Astashenok2020}). We showed that axion
field changes the behavior of scalar curvature inside and outside star in
comparison
with General Relativity. As in simple $R^2$ gravity the increase of
mass for distant observer takes place due to appearance of area
with $R\neq 0$ outside the star. But this effect is relatively
uniform for various values of density in the center of star up to
the masses close to maximum. Increase of radius also takes place
but it is not so significant. There is also some ``compensation''
between pure $R^2$ term ($\alpha R^2$) and coupling term $\beta
\phi R^2$. If $\alpha$ increases the contribution of second term
decreases due to damping mean value of curvature and axion field.

In this paper we present  rotating neutron stars in $R^2$ gravity
with axions.  We solve equations using numerical relativity's
methods and calculate characteristics of uniformly rotating
neutron stars such as mass, polar and equatorial radii and moment
of inertia. For illustration we consider realistic GM1 EoS for
neutron stars without hyperons (\citealp{GM}). This EoS is
relevant in the light of recently established sufficiently strong
limits on mass and radius for pulsar PSR J0030+0451 (see
\citealp{Riley}, \citealp{Miller}, \citealp{NICER}).

The article is organized as following. In the next section we
describe in detail the axisymmetric system of Einstein equations
in the context of $f(R)$ gravity with scalar field and methods for
solution of these equations. Then we consider star solutions
for $R^2$ gravity coupled with axion field $\phi$ in the form
$\beta R^2\phi$ where $\beta$ is some constant. Mass-radius and
mass-central energy density relations are presented for constant
frequency sequences of stars. We also considered mass-shedding
limit and calculated eccentricity and the moment of inertia for
stellar configurations. As counterparts for comparison of results
we used simple $R^2$ gravity without axions and of course General
Relativity. Results of our consideration are finally summarized
and discussed in Conclusion.

\section{3+1 formalism for rotating neutron stars in $f(R,\phi)$ gravity with
scalar field}

For $f(R,\phi)$ ($R$ is the scalar curvature) gravity with the
action
\begin{equation}\label{EinFR}
S=\frac{1}{2}\int f(R) \sqrt{-g} d^{4} x
\end{equation}
Einstein equations can be written in the form
\begin{equation}
f_{R}(R) R_{\mu\nu}-\frac{f(R)}{2} \, g_{\mu\nu} -\left(
\nabla_{\mu} \nabla_{\nu}- g_{\mu\nu}\Box\right) f_{R}(R))=8\pi
T_{\nu\mu}.
\end{equation}
Here we use system of units in which $G=c=1$. The designation
$f_R(R)$ means simply first derivative of the function $f(R)$ on
its argument $R$. Covariant D'Alambertian
$\Box=\nabla^{\mu}\nabla_{\mu}$ is introduced also. Tensor
$T_{\mu\nu}$ is the energy-momentum tensor of matter fields. For
brevity, we omit the arguments of function $f(R)$ below.

Another form of the Eq. (\ref{EinFR}) is
\begin{equation}\label{EinFR2}
f_{R}R_{\mu\nu}-\frac{1}{2}F
g_{\mu\nu}-\left(\frac{1}{2}\Box+\nabla_{\mu}
\nabla_{\nu}\right)f_{R}=8\pi
\left(T_{\mu\nu}-\frac{1}{2}g_{\mu\nu}T\right),
\end{equation}
where $T$ is the trace of energy-momentum tensor and $F\equiv
f_{R} R - f$. For description of rotation in modified gravity we
use well-known 3+1 formalism from General Relativity
(\citealp{Gourg}; \citealp{Alcubierre}; \citealp{Shapiro};
\citealp{Gourg-2}; \citealp{Sterg-13}). Let's describe
mathematical detail of this approach for $f(R)$ gravity.

Firstly one defines spacelike hypersurfaces of constant $t$ ($t$ is
coordinate time) $\Sigma_{t}$. The induced metric
$\gamma_{\alpha\beta}$ on hypersurface $\Sigma_{t}$ is
\begin{equation}
\gamma_{\alpha\beta}=g_{\alpha\beta}+n_{\alpha}n_{\beta},
\end{equation}
where $g_{\alpha\beta}$ is the metric of 4-dimensional spacetime
and $n^{\alpha}$ are components of unit timelike vector normal to
$\Sigma_{t}$. The projection operator onto hypersurface
$\Sigma_{t}$ can be defined from $\gamma{\alpha\beta}$ by raising
of the first index:
\begin{equation}
\gamma^{\alpha}_{\cdot\beta}=\delta^{\alpha}_{\cdot\beta}+n^{\alpha}
n_{\beta}.
\end{equation}

Let us consider the metric in the  following form:
\begin{equation}
ds^{2}=-N^{2}dt^{2}+\gamma_{ij}(dx^{i}+\beta^{i}dt)(dx_{j}+\beta_{j}dt).
\end{equation}
Here $N$ is so-called lapse function and  ${\beta^{i}}$ is shift
vector. Then components $n^{\alpha}$ are
$$
n^{\alpha}=N^{-1}\left(1,-\vv{\beta}\right).
$$
The next step is projecting of Einstein equations (\ref{EinFR2})
twice onto hypersurface $\Sigma_t$, twice along to $\vv{n}$ and
once on $\Sigma_t$ and along $\vv{n}$. Using well-known relations
from differential geometry one can obtain the following equations:
\begin{equation} \label{PDE1}
f_{R}\left(\der{K_{ij} }{t} - \Liec{\beta} K_{ij}\right)
      +f_{R}\left(D_i D_j N - N\left\{
      {}^3 R_{ij} + K K_{ij} -2 K_{ik} K^k_{\ \, j}\right\}\right)=
\end{equation}
$$
=4\pi N \left[ (\sigma-\epsilon) \gamma_{ij} - 2 \sigma_{ij}
\right]-\frac{1}{2}FN\gamma_{ij}-N\left(\frac{1}{2}\gamma_{ij}\Box
+D_{i}D_{j}\right)f_{R},
$$
\begin{equation}
f_{R}({}^3 R + K^2 - K_{ij} K^{ij}) = 16\pi \epsilon+F+2D^{i}
D_{i} f_{R},
   \label{PDE2}
\end{equation}
\begin{equation}
f_{R}( D_j K^j_{\ \, i} - D_i K) = 8\pi  p_i
-n^{\mu}\nabla_{\mu}(D_{i}f_{R}). \label{PDE3}
\end{equation}
In the first equation we introduced the Lie derivative of tensor of
extrinsic curvature $K_{ij}$ along the vector $\vv{n}$ \be
\label{e:ein:Lie_beta_K} \Liec{\beta} K_{ij} = \beta^k
\der{K_{ij}}{x^k}+K_{kj} \der{ \beta^k}{x^i} + K_{ik}
\der{\beta^k}{x^j}. \ee $K$ is the trace of tensor $K_{ij}$
i.e.$K=K^{i}_{i}$. 3-dimensional Ricci tensor $^{3}R_{ij}$ and
corresponding scalar curvature $^{3} R=\gamma^{ij}R_{ij}$ are
associated with the Levi-Civita connection $D$ in 3-dimensional
space. The corresponding covariant derivatives  can be expressed
through 3-dimensional Christoffel symbols ${}^3 \Gamma^i_{\ \,
jk}$ for example: \be
    D_i D_j N = \dder{N}{x^i}{x^j} - {}^3 \Gamma^k_{\ \, ij}
      \der{N}{x^k} ,
\ee
\be
D_j K^j_{\ \, i} = \der{K^j_{\ \, i}}{x^j}
      + {}^3 \Gamma^j_{\ \, jk} K^k_{\ \, i}
     - {}^3 \Gamma^k_{\ \, ji} K^j_{\ \, k} ,
\ee
\be
    D_i K = \der{K}{x^i} .
\ee The values $\epsilon$, $\sigma_{ij}$ and $p_{i}$ are defined
according to relations:
\begin{equation*}
\epsilon=n^{\mu} n^{\nu} T_{\mu\nu},
\end{equation*}
\begin{equation}
\sigma_{ij}=\gamma^{\mu}_{i}\gamma^{\nu}_{j} T_{\mu\nu},\quad
\sigma=\sigma^{i}_{i}.
\end{equation}
\begin{equation*}
p_{i}=-n^{\mu}\gamma^{\nu}_{i}T_{\mu\nu}.
\end{equation*}
and have sense of the energy density, components of stress-tensor
and vector of energy flux density correspondingly.

Now we consider star rotating along to polar axis with angular
velocity $\omega$. In this case metric can be presented in form:
\begin{equation}\label{Metr}
ds^2 = -N^2 dt^2 + A^2(dr^2 + r^2 d\theta^2) + B^2 r^2 \sin^{2}
\theta (d\phi - \omega dt)^2,
\end{equation}
where metric functions depend only on radial coordinate $r$ and
polar angle $\theta$. Shift vector $\vv{\beta}$ has  one non-zero
component:
$$
{\beta^{i}}=(0,0,-\omega).
$$
The nonzero components of extrinsic curvature for (\ref{Metr}) are
$$
K_{r\phi}=K_{\phi r}=-\frac{B^2 r^2
\sin^2\theta}{2N}\frac{\partial \omega}{\partial r}, \quad
K_{\theta\phi}=K_{\phi \theta}=-\frac{B^2 r^2
\sin^2\theta}{2N}\frac{\partial \omega}{\partial \theta}.
$$
Diagonal components are zero and therefore trace of extrinsic
curvature tensor is $K=0$. Finally one can obtain that
\begin{equation}
K_{ij} K^{ij}=\frac{B^2 r^2 \sin^{2}\theta}{2A^2
N^2}\partial\omega\partial\omega.
\end{equation}
We use the following notation
$$
\partial g_{1} \partial g_{2} \equiv \left(\frac{\partial g_{1}}{\partial
r}\frac{\partial
g_{2}}{\partial r}+\frac{1}{r^2}\frac{\partial g_{1}}{\partial
\theta}\frac{\partial g_{2}}{\partial \theta}\right).
$$
For 3-dimensional curvature we obtain
\begin{equation}
^{3}R=-\frac{2}{A^2}\left(\Delta_{(2)}\ln A + \Delta_{(4)}\ln B +
(\partial \ln B)^{2}\right).
\end{equation}
Hereinafter $\Delta_{(n)}$ is a part of Laplace operator in
$n$-dimensional Euclidean space including derivatives on radial
and polar coordinates:
$$
\Delta_{(n)}=\frac{1}{r^{n-1}}\frac{\partial}{\partial
r}\left(r^{n-1}\frac{\partial}{\partial r}\right)+\frac{1}{r^2
\sin^{n-2}\theta}\frac{\partial}{\partial
\theta}\left(\sin^{n-2}\theta \frac{\partial}{\partial
\theta}\right)
$$
The action of D'Alambertian on some function $\Phi\equiv
\Phi(r,\theta)$ depending only from radial coordinate $r$ and
polar angle $\theta$ for our task can be written as
$$
\Box \Phi = \frac{1}{A^2}\Delta_{(3)} \Phi + \frac{1}{A^2}\partial
\Phi
\partial \ln (BN).
$$
Finally for covariant Laplace operator we obtain that
$$
D^{i} D_{i} \Phi = \frac{1}{A^2} \Delta_{(3)} \Phi + \frac{1}{A^2}
\partial \Phi \partial \ln B.
$$
The trace of equation (\ref{PDE1}) gives:
\begin{equation}
f_{R}D_{i}D^{i}N=Nf_{R}({}^{3}R-2K_{ik}K^{ik})+4\pi N
(\sigma-3\epsilon)+
\end{equation}
$$
-\frac{3}{2}NF-\frac{3}{2}N\Box f_{R}-ND_{i}D^{i}f_{R}.
$$
From equation (\ref{PDE2}) it follows that
$$
f_{R}{}^{3}R=f_{R}K_{ij}K^{ij}+16\pi \epsilon+F+2D_{i}D^{i}f_{R}
$$
and therefore one can rewrite the previous equation as
\begin{equation}\label{PDE4}
f_{R}D_{i}D^{i}N=Nf_{R}K_{ij}K^{ij}+4\pi
N(\epsilon+\sigma)-\frac{1}{2}NF-
\end{equation}
$$
-\frac{3}{2}N\Box f_{R}+ND_{i}D^{i}f_{R}.
$$
Finally multiplying by $A^2/N$ and using relations for $D_{i} D^{i}$
and D'Alambertian one obtains:
\begin{equation}\label{EQ_N}
f_{R}\Delta_{(3)}\ln N + \frac{1}{2}\Delta_{(3)} f_{R}=4\pi A^2
(\epsilon+\sigma)-\frac{1}{2}A^2 F -
\end{equation}
$$
-f_{R}\partial \ln N \partial \ln(BN)-\partial \ln N
\partial f_{R} -\frac{1}{2}\partial \ln (BN)
\partial f_{R} +
$$
$$
f_{R} \frac{B^2 r^2 \sin^{2}\theta}{2N^2}(\partial \omega)^{2}
$$
Then we rewrite equation (\ref{PDE2}) using relation for
$K_{ij}K^{ij}$ and curvature $^{3}R$:
\begin{equation}\label{EQ_A}
f_{R}\Delta_{(2)}\ln A + f_{R}\Delta_{(4)}\ln B + \Delta_{(3)}
f_{R} = -8\pi A^2\epsilon -\frac{1}{2}A^2 F -
\end{equation}
$$
-f_{R} (\partial \ln B)^{2} - \partial \ln B
\partial f_{R} - f_{R}\frac{B^2 r^2 \sin^2\theta}{4N^2}(\partial
\omega)^2
$$
The next step is taking of $\phi\phi$-component of equation
(\ref{PDE1}). We use the relation for second covariant derivative
of scalar function on $\phi$
$$
D_{\phi}D_{\phi} N = \frac{B^2}{A^2} r^2 \sin^2 \theta \partial
\ln(Br\sin\theta)\partial \ln N
$$
and for $\phi\phi$-component of $3$-dimensional Ricci tensor
$$
^{3}R_{\phi\phi}=-\frac{B}{A^2}r^2 \sin^2 \theta \Delta_{(4)}B.
$$
Therefore one gets after some calculations the following
equation from (\ref{PDE1}):
\begin{equation}\label{EQ_B}
f_{R}\Delta_{(4)} \ln B + \frac{1}{2} \Delta_{(3)} f_{R}=4\pi A^2
(\sigma^{r}_{r}+\sigma^{\theta}_{\theta}-\sigma^{\phi}_{\phi}-\epsilon)-\frac{1}{2}A^2
F -
\end{equation}
$$
-f_{R} (\partial \ln B)^{2} - f_{R}
\partial\ln(Br\sin\theta)\partial \ln N - \frac{1}{2} \partial
\ln(BN)\partial f_{R}-
$$
$$
-\partial \ln(Br\sin\theta)\partial f_{R} - f_{R}\frac{B^2 r^2
\sin^{2}\theta}{2N^2}(\partial\omega)^{2}.
$$
Finally one need to get the $\phi$-component of equation
(\ref{PDE3}). Taking into account that for $R=R(r,\theta)$ and
$n^{\mu}=N^{-1}(1,0,0,\omega)$ term $n^{\mu}\nabla_{\mu}D_{\phi}
f_{R}$ is
$$
n^{\mu}\nabla_{\mu}D_{\phi}
f_{R}=f_{RR}\frac{B^2r^{2}\omega\sin^{2}\theta}{2A^2
N}\left(\partial R
\partial \ln (\omega B^4) + \frac{4}{r}\left(\frac{\partial
R}{\partial r}+\frac{1}{r\tan\theta}\frac{\partial R}{\partial
\theta}\right)\right)
$$
one obtains after multiplying by $2NA^2 B^{-2} r^{-2}
\sin^{-2}\theta$:
\begin{equation}\label{EQ_omega}
f_{R}\Delta_{(5)}\omega = -  \frac{16\pi N A^2}{B^2 r^{2} \sin^{2}
\theta}p_{\phi} +
\end{equation}
$$
+f_{RR}\left[\partial R \partial \omega + 4\omega \partial \ln B
\partial R +\frac{4\omega}{r}\left(\frac{\partial R}{\partial
r}+\frac{1}{r\tan\theta}\frac{\partial R}{\partial
\theta}\right)\right] -
$$
$$
-3 f_{R} \partial \ln B \partial \omega + f_{R} \partial \ln N
\partial \omega.
$$
One can rewrite the system of equations (\ref{EQ_N}),
(\ref{EQ_A}), (\ref{EQ_B}), (\ref{EQ_omega}). Firstly, adding
(\ref{EQ_N}) to (\ref{EQ_B}) yields
\begin{equation}\label{EQ_NB}
f_{R}\Delta_{(4)}\ln (NB) + \Delta_{(4)} f_{R} = 8\pi A^2 (
\sigma^{r}_{r} + \sigma^{\theta}_{\theta} ) - A^2 F -
\end{equation}
$$
-f_{R} (\partial \ln (NB))^{2} - 2\partial \ln(BN)\partial f_{R}.
$$
Secondly, subtracting (\ref{EQ_B}) from sum of (\ref{EQ_N}) and
(\ref{EQ_A}) gives the equation
\begin{equation}\label{EQ_NA}
f_{R}\Delta_{(2)}\ln (NA)+\Delta_{(2)} f_{R} = 8\pi A^2
\sigma^{\phi}_{\phi} - \frac{1}{2}A^2 F -
\end{equation}
$$
- f_{R} (\partial \ln N)^{2} - \partial \ln N
\partial f_{R} + \frac{3}{4}f_{R} \frac{B^2 r^2 \sin^2 \theta
}{2N^2}(\partial\omega)^{2}.
$$

For $\sigma^{\phi}_{\phi}$, $\sigma^{\theta}_{\theta}$,
$\sigma^{r}_{r}$ and

\begin{equation}
\epsilon=\Gamma^2(\rho + p) - p,
\end{equation}
\begin{equation}
\sigma^{r}_{r}=\sigma^{\theta}_{\theta}=p,\quad
\sigma^{\phi}_{\phi} = p + (\epsilon + p) U^2,
\end{equation}
\begin{equation}
p_{\phi}=B(\epsilon+p)Ur\sin\theta,
\end{equation}
where
$$
\Gamma=(1-U^2)^{-1/2}, \quad U=\frac{B}{N}(\Omega -
\omega)r\sin\theta.
$$

In $f(R)$ gravity one needs additional equation for scalar
curvature also. This equation can be obtained from trace of
Einstein equations and for our case it takes the form:
\begin{equation}\label{EQ_R}
\triangle_{(3)}f_{R}=\frac{8\pi}{3}
A^{2}(3p-\rho)-\frac{A^{2}}{3}(F-f)-\partial{\ln (NB)}\partial{
f_{R}}.
\end{equation}

For the case of function $f_{R}=F(R,\phi)$ depending also from
scalar field $\phi$ these equations are valid. For partial
derivatives of function $F(R,\phi)$  one should remember
standard rules from mathematical analysis for example
\begin{equation*}
\frac{\partial F}{\partial r}=F_{R}\frac{\partial R}{\partial
r}+F_{\phi}\frac{\partial \phi}{\partial r}
\end{equation*}
and so on.

Assuming the action for axion field in the following form
\begin{equation}
S_{\phi}=\int
d^{4}x\sqrt{-g}\left(-\frac{1}{2}\partial^{\mu}\phi\partial_{\mu}\phi-V(\phi)\right).
\end{equation}
one can obtain the equation for scalar field $\phi=\phi(r,
\theta)$:
\begin{equation}\label{scalEQ}
\triangle_{(3)}\phi=A^{2}\frac{dV}{d\phi}-\frac{A^{2}}{8\pi}\frac{df}{d\phi}-{\partial
\phi}{\partial \ln BN}.
\end{equation}

The system of equations (\ref{EQ_N}), (\ref{EQ_NA}),
(\ref{EQ_NB}), (\ref{EQ_omega}), (\ref{EQ_R}), (\ref{scalEQ})
should be supplemented by a set of boundary conditions for
functions $\nu=\ln N$, $\eta=\ln NB$, $\zeta=\ln NA$, $R$ and
$\phi$. Those are provided by the asymptotic flatness assumption.
On spatial infinity the metric tensor tends towards Minkowski
metric and therefore
$$
\nu\rightarrow 0, \quad \eta\rightarrow 0, \quad \zeta\rightarrow
0, \quad R\rightarrow 0\quad \mbox{for}\quad r\rightarrow\infty.
$$
For scalar field we also assume that $\phi\rightarrow 0$ when
$r\rightarrow +\infty$ because the density of dark matter in the
space ($\sim 10^{-29}$ g/cm$^3$) is extremely low in comparison
with densities inside relativistic stars.

For integration of system (\ref{EQ_N}), (\ref{EQ_NA}),
(\ref{EQ_NB}), (\ref{EQ_omega}), (\ref{EQ_R}), (\ref{scalEQ}) we
used the self-consistent-field method (see \citealp{Ostriker},
\citealp{Bonazzola}). The EoS is taken in the form $\rho=\rho(h)$,
$p=p(h)$, where $h$ is log-enthalpy. Zero value of $h$ corresponds
to $p=0$ (surface of star). For given central value $h_{c}$
(corresponding to some central density) the crude profile of
enthalpy chosen ($h_{c}(1-r^{2}/r_{0}^{2})$ where $r_0$ is some
radius in our calculations). Using the EoS we evaluate the pressure
and energy. Then we solve system of equations as Poisson equations
using the current $\epsilon$, $p$, $A$, $B$, $N$, $\omega$, $R$,
$\phi$ in r.h.s. of these Eqs. Therefore one can obtain the next
approximation for $A$, $B$, $N$, $\omega$, $R$ and $\phi$. Using
the useful integral of motion from Bernoulli theorem
$$
h=h_{c}+\ln N_{c}-\ln N + \ln \Gamma
$$
we get the new profile $h$. This gives new profiles of density
$\rho$ and pressure $p$. Then we again go to the solution of system as
Poisson-like equations. This procedure gives after some cycles
self-consistent solution.

\begin{figure}
\includegraphics[scale=0.36]{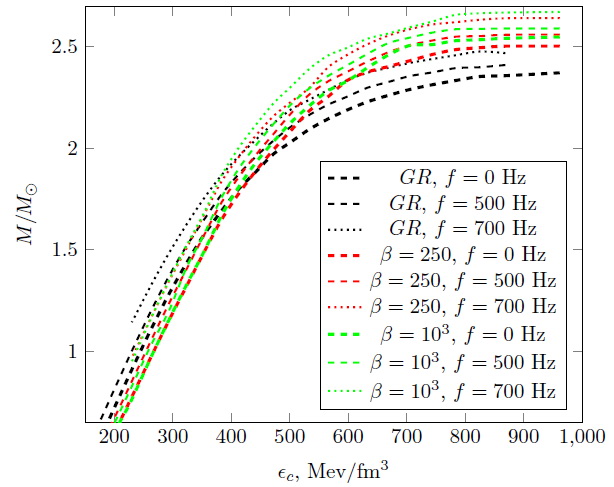}\\
\caption{Mass - central density diagram for sequences of stellar
configurations with constant frequencies of rotation (for
frequency we take $f=500$ and $700$ Hz). Black curves correspond
to non-rotating stars.}
\end{figure}

\begin{figure}
\includegraphics[scale=0.36]{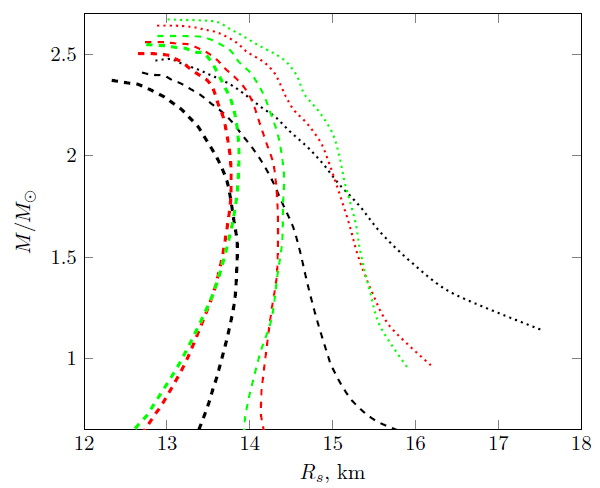}\\
\caption{Mass - equatorial radius diagram for same constant
frequency sequences as on Fig. 1. The convention of the plot
colors and symbols is the same as on Fig. 1.}
\end{figure}

\begin{figure}
\includegraphics[scale=0.36]{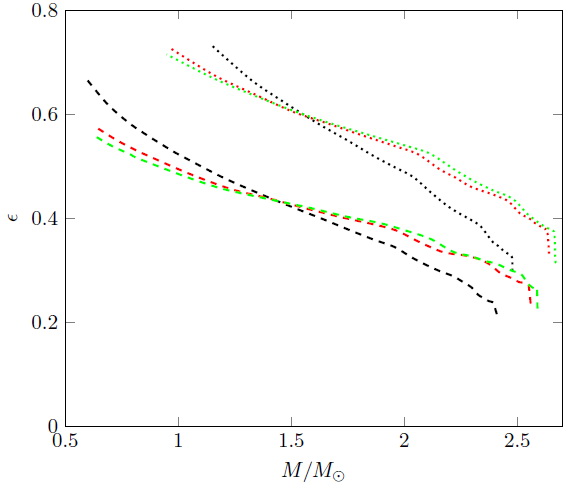}\\
\caption{Eccentricity as function of gravitational mass for
$f=500$ and $700$ Hz.}
\end{figure}

\begin{figure}
\includegraphics[scale=0.36]{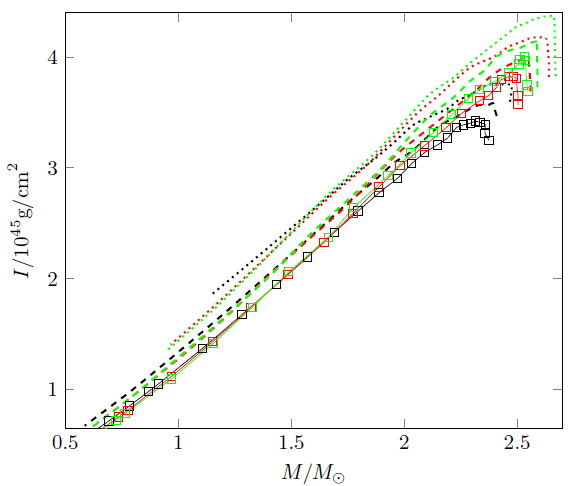}\\
\caption{Moment of inertia as function of gravitational mass for
nearly static configurations with $f=100$ (squares) and for
$f=500$ and $700$ Hz. }
\end{figure}

\begin{table}
\begin{tabular}{|c|c|c|c|}
\hline $\rho_c$, MeV/fm$^3$  &  230 & 350 & 470  \\
\hline
GR                   & 735 & 920  & 1130     \\
$\alpha=0.25$        & 745 & 950  & 1130      \\
$\alpha=2.5$        & 770 & 960 & 1145  \\
$\alpha=0.25$, $\beta=250$ & 760 & 955  & 1160 \\
$\alpha=0.25$, $\beta=1000$ & 775  & 950  & 1160   \\
\hline
\end{tabular}
\caption{Keplerian frequency (in Hz) for stellar configurations in
General Relativity, pure $R^2$ gravity for two values of $\alpha$
and for $R^2$ gravity with axions  for two values of $\beta$ (with
fixed $\alpha=0.25$) at some central densities.} \label{Table1}
\end{table}

\begin{table}
\begin{tabular}{|c|c|c|c|}
\hline Model  &  $R_{1.4}^{f=0}$ & $R_{1.4}^{f=500}$ & $R_{1.4}^{f=700}$  \\
\hline
GR                   & 13.82 & 14.67  & 16.22      \\
$\alpha=0.25$       & 13.70 & 14.47  & 15.69      \\
$\alpha=2.5$        & 13.64 & 14.33  & 15.43 \\
$\alpha=0.25$, $\beta=250$ & 13.59 & 14.33  & 15.39\\
$\alpha=0.25$, $\beta=1000$ & 13.62 & 14.33  & 15.38  \\
\hline
\end{tabular}
\caption{Radii of stellar configurations with $M=1.4M_{\odot}$ for
non-rotating case and $f=500$ and $700$ Hz in the same models as
in table 1.} \label{Table1}
\end{table}

\begin{table}
\begin{tabular}{|c|c|c|c|}
\hline Model  &  $M^{f=0}_{max}$ & $M^{f=500}_{max}$ & $M^{f=700}_{max}$  \\
\hline
GR                   & 2.39 & 2.43  & 2.49     \\
$\alpha=0.25$       & 2.44 & 2.48  & 2.53     \\
$\alpha=2.5$        & 2.51 & 2.57  & 2.62 \\
$\alpha=0.25$, $\beta=250$ & 2.47 & 2.53  & 2.58 \\
$\alpha=0.25$, $\beta=1000$ & 2.53 & 2.57  & 2.63 \\
\hline
\end{tabular}
\caption{Some parameters of neutron stars in the same models as in
Table 1: the maximal stable masses in the static case, for
frequencies $f=500$ and $700$ Hz.} \label{Table1}
\end{table}

\section{Results}

We considered in detail the model of $R^2$ gravity with the
coupling of axion field $\phi$:
\begin{equation}
F(R,\phi)=R+\alpha R^2+\beta {R^2\phi}
\end{equation}
For axion field we take simple potential without self-interaction
following \citealp{Marsh}
$$
V(\phi)=\frac{1}{2}{m^{2}_{a}}\phi^{2}.
$$
This assumption corresponds to only small deviations from the
potential minimum. For axion mass we assume the  value
corresponding to Compton wavelength $10r_{g}$ where $r_g$ is
gravitational radius of Sun ($2.95$ km). For the parameter $\beta$
the range $250<\beta<1000$ in units of $r_{g}^2$ is explored. We
also compared this model with simple $R^2$ gravity without axions
and General Relativity.

We fix two parameters (in the case of rotation) for obtaining the
stellar configuration, namely central energy density $\rho_c$ and
angular velocity $\Omega$. Varying central density in given range
we obtain sequence of neutron stars rotating with constant angular
velocity.

Asymptotical behavior of $A(r, \theta)$ at $r\rightarrow\infty$
defines gravitational mass $M_s$ of star for distant observer. In
General Relativity the solution of Einstein equations outside the
star has the form for non-rotating star:
\begin{equation}
A(r)=\left(1+\frac{M_s}{2r}\right)^{2}, \quad
N(r)=\left(1-\frac{M_s}{2r}\right)\left(1+\frac{M_s}{2r}\right)^{-1}.
\end{equation}

Therefore, the gravitational mass of star can be found  as an
asymptotical limit
\begin{equation*}
M_s=2\lim_{r\rightarrow\infty}r(\sqrt{A}-1).
\end{equation*}
One should also account that circumferential radial coordinate
${r_c}$ is
\begin{equation*}
\tilde{r_c}=Ar.
\end{equation*}
Note that in the following the symbol ``r'' on figures means
circumferential radius. Suffix ``c'' is omitted.

On figure 1 we depicted the gravitational mass-central energy
density relations for static case and three values of rotation
frequency (for our calculations we considered the cases of
$f=100$, $500$ and $700$ Hz). From these diagrams one can see that
gravitational mass of star is increased with rotation in our models
as in General Relativity.

The mass-equatorial radius diagram for same $f$-constant sequences
are depicted on Fig. 2. One need to point out that parameters of
stars weakly depend from frequency up to $f\sim 200$ Hz. The same
limit is well-known in General Relativity. The interesting problem
is to find the limits on masses of fast rotating neutron stars.
From observations as well-known fastest rotating pulsar is PSR
J1748 - 2446ad with $f=716$ Hz (\citealp{FAST}). The minimum mass
of a rotating neutron star depends from frequency of rotation and
chosen EoS. We can estimate the lower mass of stars with $f=700$
Hz in our model in comparison with General Relativity. From
calculations it follows that lower bound for fast rotating neutron
stars in our models can be considerably reduced for given EoS. For
$f=700$ Hz this limit for $\alpha=0.25$, $\beta=250$ is only $\sim
M_\odot$ (in General Relativity we have $\sim 1.2M_{\odot}$). For
another EoS choice one can expect the same picture: in $R^2$
gravity with axions fast rotating stars with smaller masses can
exist. From results of \citealp{Zdunic} and \citealp{Cipoletto} it
follows that in General Relativity the minimal neutron star masses
at $f=716$ Hz for another stiff EoS (NL3, TM1) lie in the range
$1.3-1.4 M_{\odot}$. In principle the possible future observation
of a fast rotating neutron stars with a lower mass could rule out
these EoS or could be considered as some indirect confirmation of
alternative models of gravity.

Second feature of stellar configurations  in $R^2$ gravity with
axions is that stars with intermediate masses
($1.1M_{\odot}<M<1.4M_{\odot}$) are more compact in comparison
with General Relativity (see Table 2). The difference between radii of stars at
$700$ Hz is $\sim 1$ km for $M=1.4M_\odot$ and up to $2$ km for
$M=1.2M_\odot$. Therefore in modified gravity one can expect the appearance of
more compact fast rotating stars. It is noteworthy
that such difference can be tested from observations. X-ray
astronomy allows to determine radii of neutron stars with better
precision. This effect take place only for fast rotating stars.
From Fig. 2 one can see that for static stars the difference
between radii is negligible for given interval of masses.

The next question is maximal possible frequency of rotation in our
models. Sequence of stellar configurations with given central
density  will end up at so-called Keplerian frequency $\Omega_{k}$.
Hydrostatic equilibrium does not exist for star with
$\Omega>\Omega_{k}$ because gravitational force exceeds the
centrifugal force at the equator and therefore expulsion of mass
from the star begins. This mass-shedding limit on frequency for
our model of gravity is higher in comparison with General
Relativity. In Table 1 we give the values of Keplerian frequency
for some central densities.

In Table 3 we give for comparison some parameters of neutron stars
in General Relativity, pure $R^2$ gravity (for two values of
$\alpha$) and  $R^2$ gravity with axions  for two values of
$\beta$ (with fixed $\alpha=0.25$): the maximal stable mass in the
static case and for frequencies $f=500$ and $700$ Hz.

Maximal mass of star increases with rotation as expected. Effect of increasing
mass due to the coupling between axion field and scalar curvature also is
observed. For maximal mass of star with $f=700$ Hz in a case of GM1 EoS we
obtained the value $\sim 2.65M_{\odot}$ for $\alpha=0.25$ and $\beta=10^3$ in
comparison with
$M_{max}=2.47M_{\odot}$) in General Relativity. In opposite  case of
non-rotating stars when difference between radii of stars in modified gravity
and General Relativity is $\sim 0.4$ km for $f=500$ and $700$ Hz this quantity
is reduced. This occurs because in General Relativity radius of star increases
with rotation more strongly in comparison with our model of gravity.

The next interesting question is to investigate the deformation of
star caused by rotation. We calculated the eccentricity parameter
for stellar configurations as
\begin{equation}
e=\left(1-\frac{R_{p}^2}{R_{eq}^2}\right)^{1/2}
\end{equation}
From Fig. 3 one can see that forms of stellar configurations in
our model and in General Relativity in principle are similar for
corresponding frequencies.

In description of pulsar properties the main important quantity is
the moment of inertia. It can be calculated as
\begin{equation}
     I=\frac{J}{\Omega},
\end{equation}
where $J$ is angular momentum. For calculation of angular momentum we used the
relation from General Relativity namely
\begin{equation}
     J=\int_{\Sigma_t} (\epsilon+p)UA^2 B^2 r^3 \sin^2\theta dr d\theta d\phi.
\end{equation}
This approximation can be considered as realistic because from the
physical viewpoint the inertial characteristics of neutron stars
should  depend only from solution inside star unlike gravitational
mass. Moment of inertia cannot be obtained directly from
observations.  We depict the moment of inertia as a function of
the gravitational mass for frequency constant sequences on Fig. 4
for our model in comparison with General Relativity. Only for
masses close to maximal one the inertial moment in modified
gravity considerably declines from corresponding value in General
Relativity for same frequency. This deviation can affect in
principle the evolution of the spin period of massive fast
rotating pulsars. Unfortunately we have no a lot of observational
data about such pulsars.

One should also note that we considered only one value for axion
mass $m_{a}=0.1$ (in units of $r_g^{-1}$. In case of non-rotating
stars for smaller masses (for example in previous work we
considered $m_{a}=0.01$) only size of axion ``galo'' around the
star increases but scalar curvature for $r>50$ km is very close to
zero and therefore contribution of term $R\phi^2$ is negligible.
For stars rotating with frequencies up to $700$ Hz for $r>\sim 20$
km the solution of Einstein equations is very close to static
spacetime. Therefore, the rotation parameters of stars weakly
depend from parameter $m_{a}$ (of course for $\lambda_c>\sim
R_s$).

\section{Conclusion}

We investigated realistic model of a uniformly rotating neutron
star in axion $R^2$ gravity with curvature-axion coupling in the
form $\sim R^2\phi$. For description of nuclear matter GM1 EoS is
used. We calculated gravitational mass, equatorial and polar
radii, eccentricity and moment of inertia for stellar
configurations with constant frequency.

As in non-rotating case the increase of stellar mass due to axion
scalar field takes place for rotating stars and in principle this
effect weakly depends from frequency of rotation. One notes also
that star radius for our model increases but not significantly for
fast rotation. We obtained the increase of mass $\sim
0.2M_{\odot}$ for massive stars in the case of $\beta=1000$. This
value is sufficient for possible observational indication of such
model. {The star radius increases not so considerably ($\sim 100$
m for $\beta=1000$).

Maximal possible frequency of rotation (Keplerian or mass-shedding limit) in
$R^2$ gravity with
axion increases in comparison with General Relativity. Of course
this fact  is interesting only from theoretical point of view
because we have no observational data about compact stars rotating with
frequency closed to mass-shedding limit.

However, our results show another interesting (and in principle observable in
future) features of stellar configurations
in modified gravity with axions. Stars with intermediate masses
$M_{\odot}<M<1.4
M_{\odot}$ are more compact at the same frequency of rotation.
This difference for some parameters ($\sim 1-2$ km) in principle
lies in possible errors of radii measurement from NICER mission.
Neutron stars in modified gravity in some sense are more stable to fast
rotation and mass bound on fast rotating pulsars become lower (for fixed EoS of
course). We obtained for GM1 EoS that the limit on mass of fastest rotating
neutron star (with $f\sim 700$ Hz) is close to $M_{\odot}$ down  $20\%$ than in
General Relativity for this EoS. Our preliminary calculations for another
realistic EoS (Sly4 and APR) give the similar results. Possible observation of
fast, compact stars with relatively small masses could be the best proof of
viability of current model of gravity with axions.}

Analysis of recently observed GW event (\citealp{Abbott})
indicates towards possible existence of neutron stars with mass
$2.7 M_\odot$. This upper limit (if it is reliably confirmed) in
combination with constraints on neutron star radii from NICER puts
the question about validity of many realistic equations of state
for dense matter for example even GM1 EoS considered in paper. In
frames of model of $R^2$ gravity with axion it is possible to get
the increase of observed neutron star masses for required limit
for this EoS. It is interesting to consider another modified
gravities as well as different interactions between gravity and
axions which can lead to increase of maximal mass. We plan to
consider this in near future.

\textbf{Data availability.} No new data were generated or analysed
in support of this research.

\bibliographystyle{mnras}

\label{lastpage}
\end{document}